\documentclass[aps,prl,superscriptaddress]{revtex4-2}

\usepackage{amsmath}
\usepackage{graphicx}
\usepackage{MnSymbol}
\usepackage{parskip}
\usepackage{makecell}
\usepackage{booktabs}
\usepackage{multirow}
\usepackage{xcolor}

\begin{document}

\title{On the flash temperature in sliding rubber contacts}

\author{B.N.J. Persson}
\affiliation{Peter Gr\"unberg Institute (PGI-1), Forschungszentrum J\"ulich, 52425, J\"ulich, Germany}
\affiliation{State Key Laboratory of Solid Lubrication, Lanzhou Institute of Chemical Physics, Chinese Academy of Sciences, 730000 Lanzhou, China}
\affiliation{MultiscaleConsulting, Wolfshovener str. 2, 52428 J\"ulich, Germany}

\begin{abstract}
We present an analytical theory for the flash temperature for viscoelastic solids sliding on
rigid and randomly rough surfaces. The theory takes into account the surface roughness on all relevant length scales. 
\end{abstract}

\maketitle

\setcounter{page}{1}
\pagenumbering{arabic}

%\pagestyle{empty}

%%%%%%%%%%%%%% main text %%%%%%%%%%%%%%%%
%\begin{multicols}{2}

%%%%%%%%%%%%%% main text %%%%%%%%%%%%%%%%

{\bf 1 Introduction}

Friction between surfaces generates heat, leading to temperature increases at the contact points.
This phenomenon is known as flash temperature, which is the high, localized, and brief temperature spike that
occurs at the true points of contact between two rubbing solids. The rapid generation of heat at these locations
causes thermal spikes, resulting in intense flash temperatures as kinetic energy is converted into
heat. These spikes can be extremely high, sometimes reaching
over $1000\,^\circ {\rm C}$, but they are also incredibly brief,
lasting only for the instant that the asperities are in contact.
This process is so rapid that the generated heat has little time to conduct away into the bulk of the materials,
trapping thermal energy and further elevating the temperature at the contact points.

In almost all cases, most of the dissipated energy in sliding friction ends up as thermal energy within the sliding 
solids. The temperature field in the solids can be written as $T({\bf x},t) = T_0({\bf x},t) + \Delta T({\bf x},t)$.
The {\it background} temperature $T_0({\bf x},t)$ varies slowly in space and time while the {\it flash} temperature
$\Delta T({\bf x},t)$, varies rapidly in space and time. 
$\Delta T({\bf x},t)$ is non-zero only close to the asperity contact regions and is highly localized in space.

Frictional heating is important in a wide range of applications, spanning from ice and rubber friction to the
sliding of minerals like granite, which is central to earthquake dynamics.  The flash temperature
occurs in the asperity contact regions, where frictional energy is converted into heat. This localized heating
can have a crucial influence on the resulting friction, usually reducing it. This is the case for sliding ice,
where the flash temperature can melt the ice \cite{ice1,ice2,ice3}, or at least shift the temperature in the contact region towards the
melting point \cite{ice4,ice5}, thereby reducing the frictional shear stress and the friction force.
The same is true for granite
sliding on granite, where the melting point of the mineral (primarily quartz) may be reached at the sliding speeds (on the order of $\sim 1 \ {\rm m/s}$)
involved in earthquakes \cite{earth1,earth2,earth3}. Rubber friction depends exponentially on the temperature, and an increase in temperature
shifts the friction coefficient master curve to higher sliding speeds, which usually reduces the friction but sometimes
increases it \cite{rub1,rub2,rub3,Miy}.

\begin{figure}
\includegraphics[width=0.27\textwidth,angle=0.0]{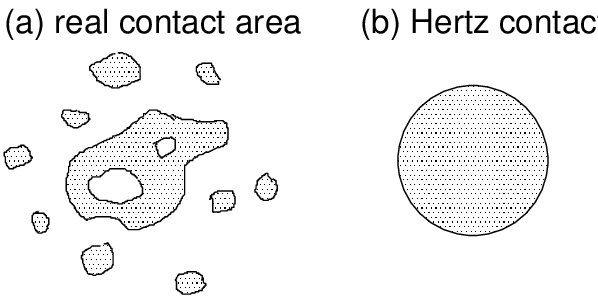}
\caption{\label{MacroAsperityTopSide.eps}
(a) A macroasperity contact area and (b) the contact are in Hertz approximation.
}
\end{figure}

In Ref. \cite{MP} we have present a multiscale theory for the flash temperature during steady sliding.
The theory was based on the study of  temperature-temperature and temperature-stress correlation functions and included all relevant length scales.
In the limiting case of roughness on a single length scale, the theory reduces to the classical
theories of Jaeger, Archard, and Greenwood \cite{flash1,flash2,flash3} (see also \cite{flash4,flash5,flash6,TianGreen,ReddyhoffHardening,ZhuNumerical}).
These authors studied the temperature resulting from moving heat sources with circular shape with constant or Hertz-like pressure profiles. 
However, in reality the asperity contact regions have complex shapes where the macroasperity contact regions
consist of agglomerates of much smaller contact regions (see Fig. \ref{MacroAsperityTopSide.eps}) with complex
internal pressure distributions. In Ref. \cite{MP} it was shown
that for real surfaces with multiscale roughness the classical theories for the flash temperature fail severely.
The theory was compared to experimental data for steel sliding on steel in Ref. \cite{P1}.

In the following section (Sec. 2) all temperatures refer to the {\it increase} in the temperature above the background temperature $T_0(t)$.
Thus $T_q$ is a weighted average flash temperature,
and the actual (average) temperature in the contact regions is $T_0(t)+T_q(t)$. 
Similarly, the temperature-stress correlation function $\langle T({\bf x},t)\sigma ({\bf x},t)\rangle$ is calculated with the background temperature
$T_0$ subtracted from $T({\bf x})$. 
Stated differently, all temperatures refer to actual temperatures only if the background temperature vanishes.

Temperature effects are particularly important for rubber friction.
In earlier studies a theory was developed for rubber friction including the flash temperature
\cite{rub1,rub2,rub3}. However, the flash temperature
was calculated by smearing out the frictional power density laterally within the macroasperity contact regions, 
an approach similar to the classical one length-scale theories for the flash temperature.
Here we will generalize the theory developed in Ref. \cite{MP} to viscoelastic solids, focusing
on rubber sliding on rigid randomly rough surfaces. 

\vskip 0.2cm
{\bf 2 Theory}

For a rubber block sliding on a hard randomly rough surface there are two contribution to the friction coefficient,
one from the viscoelastic deformations of the rubber by the substrate asperities 
on many length scales denoted $\mu_{\rm visc}$, and another from the adhesive interaction between the 
rubber molecules and the substrate in the area of real contact denoted $\mu_{\rm ad}$. 
Both $\mu_{\rm visc}$ and $\mu_{\rm ad}$ depends sensitively on the temperature. 

In what follows we will denote wavenumber by $q$ and sometimes, when occurring as dummy variables in integrals, also by $k$ and $s$.
The viscoelastic contribution to the friction coefficient can be written as\cite{P2,P3} 
$$\mu_{\rm visc} \approx {1\over 2 \sigma_0} \int_{q_0}^{q_1} dq \ q^3 C(q) P(q) S(q) \int_0^{2\pi} d\phi \ {\rm cos}\phi \, {\rm Im} E^*(qv \, {\rm cos} \phi , T_q)\eqno(1)$$
where $\sigma_0$ is the applied nominal stress and $T_q$ an effective temperature defined below.
In (1) $E^* (\omega,T) = E(\omega, T)/(1-\nu^2)$ is the effective viscoelastic modulus for the frequency 
$\omega = qv \, {\rm cos} \phi $ and the effective temperature $T=T_q$, and
$C(q)$ is the surface roughness power spectrum, and 
$$S (q) = \gamma + (1-\gamma)P^2,\eqno(2)$$
where
$$P(q)={\rm erf}\left ({\sigma_0 \over 2 \surd G}\right )\eqno(3)$$
where
$$G(q)= {1\over 8} \int_{q_0}^q ds \ s^3 C(s) \int_0^{2\pi} d\phi \, |E^* (sv \, {\rm cos} \phi, T_s)|^2, \eqno(4)$$
The function $P(q) = A(q)/A_0$ is the relative contact area observed when the interface is studied at the magnification $\zeta = q/q_0$.

The contribution to the friction force from the area of real contact is assumed proportional to the area of real contact $A$
and given by $\tau A$, where $\tau$ is the frictional shear stress acting in the area of real contact which depends on the sliding speed and the temperature
$T_{\rm R}$ at the rubber surface in the area of real contact. The corresponding friction coefficient
$$\mu_{\rm ad} = {\tau A \over \sigma_0 A_0} = {\tau \over \sigma_0} P(q_{a})$$
where $q_a \sim 1/a$, where $a$ is an atomic or molecular length.

Here we will derive the effective temperatures $T_q$ and $T_{\rm R}$ to be used when calculating the friction coefficients.
We assume a constant sliding speed $v$ and neglect heat transfer to the substrate.

When the interface is studied at the magnification $\zeta$ it appears that the
surfaces makes contact in the area $A(\zeta)$. The normal apparent stress
acting in the contact area is $\sigma ({\bf x},\zeta)$ and is non-vanishing only
where the surfaces makes (apparent) contact when observed at the magnification $\zeta$. 
Note that for any magnification the contact stress $\sigma ({\bf x},\zeta)$ 
integrated over the surface area must equal the normal load $\sigma_0 A_0$:
$$\int d^2 x \ \sigma ({\bf x},\zeta) = \sigma_0 A_0 \eqno(5)$$

Let $f'(\zeta) d \zeta$ be the contribution to the
viscoelastic friction power from the roughness components which appear when the magnification increases from
$\zeta$ to $\zeta+d\zeta$. The viscoelastic contribution to the
dissipated energy per unit time and unit volume from
the roughness components between $\zeta$ to $\zeta+d\zeta$ is written as
$$d \dot q({\bf x},z,\zeta) =\dot q({\bf x},z,\zeta+d\zeta) -\dot q({\bf x},z,\zeta) = 
\sigma ({\bf x},\zeta) g(z,\zeta) f'(\zeta) d\zeta \eqno(6)$$
where
$$g(z,\zeta) = 2 q_0 \zeta e^{-2 q_0 \zeta z} \eqno(7)$$
is the dependency of $ d \dot q$ on the distance $z$ into the solid. 
Here we have used that a surface stress component $\sim {\rm exp} (i {\bf q}\cdot {\bf x})$ generate
a deformation field which decay into the solid like $\sim {\rm exp}(-qz)$ and since the dissipated energy
scales like the square of the strain we expect $\dot q \sim {\rm exp}(-2qz)$. 
Note the normalization
$$\int_0^\infty dz \ g(z,\zeta) = 1 . \eqno(8)$$

Using (6) we get
or
$${d \dot q \over d \zeta} = \sigma ({\bf x},\zeta) g(z,\zeta) f'(\zeta) $$
or
$$\dot q({\bf x}, z, \zeta) = \int_1^\zeta d\zeta'  \ \sigma ({\bf x},\zeta') g(z,\zeta' )  f'(\zeta') . \eqno(9)$$

Integrating (9) over the volume of the solid and using (5) and (8) gives
$$\int d^2 x \int_0^\infty dz \ \dot q({\bf x},z, \zeta ) =  \int_1^\zeta d\zeta' \  f'(\zeta') \sigma_0 A_0 =f(\zeta) \sigma_0 A_0 $$
which must equal $\mu_{\rm visc} (\zeta ) \sigma_0 A_0 v$ where $\mu_{\rm visc} (\zeta)$ is the friction coefficient
including the roughness components $q < \zeta q_0$, observed when the magnification equals $\zeta$. Hence we get
$$f(\zeta) = \mu_{\rm visc} (\zeta) v \eqno(10)$$

Assuming that $\zeta_1 $ is the highest magnification, where all the roughness components with the wavenumbers $q<q_1 = \zeta_1 q_0$ is included, we 
obtain the total heat energy density
$$\dot q ({\bf x},z) = \dot q({\bf x}, z, \zeta_1) =\int_1^{\zeta_1} d\zeta  \ h(\zeta) \sigma ({\bf x},\zeta) e^{-2 q_0 \zeta z} \eqno(11)$$
where
$$h(\zeta) =  2 q_0 \zeta v \mu_{\rm visc}' (\zeta ) . \eqno(12)$$

The temperature distribution in a sliding contact is considered in the steady state after run-in.
We analyze the heat diffusion in the half-space $z>0$ with a heat source (energy per unit volume and unit time)
$\dot q ({\bf x},z,t) = \dot q (x,y,z,t)$. We will first assume that the heat current $J_z$ vanishing on the surface $z=0$. This
heat diffusion problem can be solved most easily by extending the problem to the heat diffusion
in an infinite solid with a heat source which is symmetric $\dot q ({\bf x},z,t)=\dot q ({\bf x},-z,t)$.
Thus, using (11) and writing
$$\sigma ({\bf x},\zeta) = \int d^2q \, \sigma ({\bf q},\zeta) e^{i{\bf q} \cdot {\bf x}}\eqno(13)$$
and
$$\sigma ({\bf q},\zeta) = {1\over (2 \pi )^2} \int d^2x \ \sigma ({\bf x},\zeta) e^{-i {\bf q} \cdot {\bf x}} \eqno(14)$$ 
we get
$$\dot q ({\bf x},z) =\int_1^{\zeta_1} d\zeta  \, h(\zeta) \int d^2q \, \sigma ({\bf q},\zeta) e^{i{\bf q}\cdot {\bf x}-2 q_0 \zeta |z|} \eqno(15)$$
Using that
$$e^{-2q_0 \zeta |z|} = {1\over 2 \pi} \int dk_z {4 (q_0 \zeta) \over k_z^2 + 4 (q_0 \zeta)^2} e^{ik_z z}$$
we get
$$\dot q ({\bf x},z) = {1\over 2 \pi} \int_1^{\zeta_1} d\zeta  \ h(\zeta) \int dk_z {4 (q_0 \zeta) \over k_z^2 + 4 (q_0 \zeta)^2}    
\int d^2q \ \sigma ({\bf q},\zeta) e^{i [{\bf q} \cdot {\bf x}+k_z z]} \eqno(16)$$

The temperature distribution in the sliding solid is determined by the heat diffusion equation
$$\rho c_{\rm P} {\partial T \over \partial t} -  \kappa \nabla^2 T  = \dot q ({\bf x}, z,t) \eqno(17)$$
where $\rho$, $c_{\rm P}$, and $\kappa$ are the mass density, the specific heat capacity, and the thermal conductivity,
respectively. Introducing the thermal diffusivity $D=\kappa/\rho c_{\rm P}$, we can write
$${\partial T \over \partial t} -  D \nabla^2 T  = {D \over \kappa } \dot q ({\bf x},z,t), \eqno(18)$$
where $\dot q $ depends on the position $({\bf x},z) =(x,y,z)$ and time $t$.

We consider first sliding at a constant speed ${\bf v}$ so that the heat source
$$\dot q ({\bf x},z,t) = \dot q({\bf x}-{\bf v} t,z)$$ 
Using (16) and (18) gives
$$T ({\bf x},z,t) = {1\over  2 \pi \kappa } \int_1^{\zeta_1} d\zeta  \ h(\zeta) \int dk_z {4 (q_0 \zeta) \over k_z^2 + 4 (q_0 \zeta)^2} \int d^2q \, {  \sigma  ({\bf q}, \zeta) \over q^2+k_z^2- i{\bf q}\cdot {\bf v}/D} e^{i[{\bf q} \cdot ({\bf x}-{\bf v}t)+k_z z]}. \eqno(19)$$
Changing the integration variable ${\bf q}$ to ${\bf k}$ and defining
$$\sigma  ({\bf k}, t, \zeta) = \sigma  ({\bf k},\zeta) e^{-i {\bf k} \cdot {\bf v} t} \eqno(21)$$ 
we get
$$T ({\bf x},z,t) = {1\over  2 \pi \kappa } \int_1^{\zeta_1} d\zeta  \ h(\zeta) \int dk_z {4 (q_0 \zeta) \over k_z^2 + 4 (q_0 \zeta)^2} \int d^2k \, {  \sigma  ({\bf k}, t, \zeta) \over k^2+k_z^2- i{\bf k}\cdot {\bf v}/D} e^{i[{\bf k} \cdot {\bf x}+k_z z]}. \eqno(20)$$

This temperature is non-vanishing everywhere in the solid but we are only interested in the temperature
increase in the asperity contact regions where the frictional energy is produced. This is mainly the regions close to the surface where
the contact stress $\sigma ({\bf x},t)$ in non-vanishing. 
When calculating the contribution to the rubber friction the roughness wavevector component ${\bf q}$ will give a contribution to the heat source which decay roughly as
${\rm exp} (-2qz)$ into the solid, where $q=|{\bf q}|$. The effective temperature $T_q$ to use in the expression for the friction coefficient can be obtained by
``projecting'' $T({\bf x}, z, t)$ on  $\sigma ({\bf x},t,\zeta) e^{-2qz}$. Thus we define 
$$T_q (t) =  {\langle T ({\bf x},z,t) \sigma ({\bf x},t) e^{-2qz}\rangle \over \langle \sigma ({\bf x},t) e^{-2qz}\rangle } \eqno(23)$$
We have
$$\langle \sigma ({\bf x},t) e^{-2qz}\rangle = 
\int d^2x \, \sigma ({\bf x}-{\bf v}t) \int_0^\infty dz \, e^{-2qz} = {\sigma_0 A_0  \over 2q} \eqno(24)$$
Using that
$$\int d^2x \int_0^\infty dz \, \sigma ({\bf x},t) e^{-2qz} e^{i[{\bf k} \cdot {\bf x}+k_z z]} = (2\pi )^2 {\sigma (-{\bf k},t) \over 2q-ik_z}$$
we get
$$\langle T ({\bf x},z,t) \sigma ({\bf x},t) e^{-2qz}\rangle = {1\over  2 \pi \kappa } \int_1^{\zeta_1} d\zeta  \ h(\zeta) \int dk_z {4 (q_0 \zeta) \over k_z^2 + 4 (q_0 \zeta)^2} 
(2\pi )^2 \int d^2k \, {\langle \sigma  ({\bf k}, t, \zeta) \sigma (-{\bf k},t) \rangle \over (k^2+k_z^2- i{\bf k}\cdot {\bf v}/D) (2q-ik_z)}. \eqno(25)$$
Since
$$\langle \sigma  ({\bf k}, t, \zeta) \sigma  (-{\bf k}, t) \rangle = \langle \sigma  ({\bf k}, \zeta) e^{i{\bf k}\cdot {\bf v t}} \sigma  (-{\bf k}) e^{-i{\bf k}\cdot {\bf v t}} \rangle 
= \langle \sigma  ({\bf k}, \zeta) \sigma  (-{\bf k}) \rangle$$
(25) is, as expected, time-independent.

The stress power spectrum \cite{Persson1}
$$\langle \sigma ({\bf k}) \sigma (-{\bf k})\rangle = {A_0\over (4\pi)^2} |E^*(\omega,T)|^2  k^2 C(k) W(k), \eqno(26)$$
where $E^* (\omega,T) = E(\omega,T)/(1-\nu^2)$, with $\omega = {\bf k}\cdot {\bf v}$ and $T=T_k$, is the effective viscoelastic modulus,
$C(k)$ is the surface roughness power spectrum, and where $W(k)$, $P(k)$ and $G(k)$ obtained from (2)-(4) with $q$ replaced by $k$.
The stress power spectrum $\langle \sigma ({\bf k},\zeta) \sigma (-{\bf k})\rangle$ equals (26) for $k < \zeta q_0$, but vanish for
$k > \zeta q_0$ because at the magnification $\zeta$ only the roughness components with wavenumber $k < \zeta q_0$ can be observed.
Using this result and (24) and (26) in (25) gives
$$T_q = {1\over  8 \pi \kappa \sigma_0} \int_1^{\zeta_1} d\zeta  \ 2 q_0 \zeta v \mu_{\rm visc}'(\zeta) $$
$$\times \int_{q_0}^{q_0\zeta} dk \, k^3 C(k) W(k) \int_0^{2\pi} d\phi \ 
 \int dk_z {4 (q_0 \zeta) 2q |E^*(vk {\rm cos}\phi , T_k)|^2    \over [k_z^2 + 4 (q_0 \zeta)^2] [2q-i k_z] [k^2+k_z^2 - i (kv/D) {\rm cos}\phi ] }. $$
If we denote $q_0 \zeta = s$ we get the viscoelastic contribution to the flash temperature
$$T_q ({\rm visc})= {2 v\over  \pi \kappa \sigma_0} \int_1^{q_1} ds  \ \mu_{\rm visc}'(s) $$
$$\times \int_{q_0}^{s} dk \, k^3 C(k) W(k) \int_0^{2\pi} d\phi \ 
 \int_{-\infty}^{\infty} dk_z {s^2 q |E^*(vk {\rm cos}\phi ,T_k)|^2  \over (k_z^2 + 4 s^2)(2q-i k_z)[k^2+k_z^2 - i (kv/D) {\rm cos}\phi ]}. \eqno(27)$$
From (1) we have
$$\mu_{\rm visc}'(q) =  {q^3 \over 2 \sigma_0} C(q) P(q) S(q) \int_0^{2\pi} d\phi \ {\rm cos}\phi \, {\rm Im} E^*(qv \, {\rm cos} \phi , T_q)  \eqno(28)$$
The $k_z$-integral in (27) can be performed analytically using that
$$\int_{-\infty}^\infty dk_z {1 \over (k_z^2 +a^2) (b-ik_z) (k_z^2+c^2)} = {\pi \over (c^2-a^2)}\left [{1\over a(a+b)}-{1\over c (c+b)}\right ]$$

Eq. (27) can also be used for the adhesive contribution to the temperature from the area of real contact. 
The heat source associated with the adhesive friction is localized to molecular thin ( $\sim 1 \ {\rm nm}$) layer 
at the rubber surface in the area of real contact so we can assume that it decay into the solid
like ${\rm exp}(-2 q_{\rm a} z)$ where $q_{a} = 1/a$ where $a$ is a molecular length. In the limit $q_a \rightarrow \infty$ this corresponds to using
$$\mu_{\rm ad}' (q) = \mu_{\rm ad} \delta (q-q_a)$$
Thus if the cut-off wavenumber $q_1 > q_{a}$ we have
$$\int_1^{q_1} dq  \ \mu_{\rm ad}'(q) = \mu_{\rm ad}\eqno(29)$$
and the contribution to the temperature from the adhesive frictional interaction becomes
$$T_q = {2 v \mu_{\rm ad} \over  \pi \kappa \sigma_0} \int_{q_0}^{q_{1}} dk \, k^3 C(k) W(k)\int_0^{2\pi} d\phi \ 
 \int_{-\infty}^{\infty} dk_z {q_{a}^2 q |E^*(vk {\rm cos}\phi , T_k)|^2  \over (k_z^2 + 4 q_{a}^2)(2q-i k_z)[k^2+k_z^2 - i (kv/D) {\rm cos}\phi ]}.$$
In this equation one can take the limit $q_a \rightarrow \infty$ to get
$$T_q ({\rm ad})= {v \mu_{\rm ad} \over  2 \pi \kappa \sigma_0} \int_{q_0}^{q_{1}} dk \, k^3 C(k) W(k)\int_0^{2\pi} d\phi \ 
 \int_{-\infty}^{\infty} dk_z {q |E^*(vk {\rm cos}\phi , T_k)|^2  \over (2q-i k_z)[k^2+k_z^2 - i (kv/D) {\rm cos}\phi ]}. \eqno(30)$$
The limit $q_a \rightarrow \infty$ correspond to assuming that the adhesive contribution to the flash temperature arising from a heat source of the form
$$\dot q ({\bf x},z) = \mu_{\rm ad} v \sigma ({\bf x}) \delta (z) \eqno(31)$$
The total flash temperature is given by the sum of (27) and (30) and the flash temperature $T_{\rm R}$ at the surface $z=0$ can be obtained from this sum 
in the limit $q \rightarrow \infty$.

As an illustration, the flash temperature at the surface $z=0$ from the adhesive contribution 
can be obtained from (30) by taking the limit $q \rightarrow \infty$ 
$$T_\infty = {v \mu_{\rm ad} \over  4 \pi \kappa \sigma_0} \int_{q_0}^{q_{1}} dk \, k^3 C(k) W(k)\int_0^{2\pi} d\phi \ 
 \int_{-\infty}^{\infty} dk_z { |E^*(vk {\rm cos}\phi , T_k)|^2  \over [k^2+k_z^2 - i (kv/D) {\rm cos}\phi ]}$$
$$ = {v \mu_{\rm ad} \over  \kappa \sigma_0} \int_{q_0}^{q_{1}} dk \, k^2 C(k) W(k) 
{\rm Re} \int_0^{\pi /2} d\phi \ {|E^*(vk {\rm cos}\phi , T_k)|^2 \over [1-i(v/Dk)^2 {\rm cos}\phi ]^{1/2}}. \eqno(32)$$

The flash temperature at the surface for elastic contact with the heat source at the surface $\dot q ({\bf x},t) \sim \delta (z)$ was studied in Ref. \cite{M1} and can be obtained
from (30) assuming $E^*$ a constant. Thus in this limit (32) reduces to
$$T_{\rm flash} = {v \mu_{\rm ad} \over  \kappa \sigma_0} (E^*)^2 \int_{q_0}^{q_{1}} dk \, k^2 C(k) W(k) {\rm Re} \int_0^{\pi /2} d\phi \ {1\over [1-i(v/Dk)^2 {\rm cos}\phi ]^{1/2}}.$$
which agree with the result obtained in Ref. \cite{MP}.

\begin{figure}
\includegraphics[width=0.25\textwidth,angle=0.0]{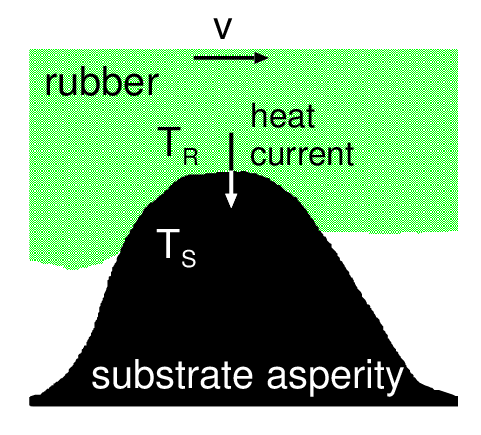}
\caption{\label{asperity.eps}
A heat current will flow from the rubber to the substrate in the area of contact.
}
\end{figure}

\vskip 0.2cm
{\bf 3 Heat transfer}

When a rubber block is slid on a substrate heat energy will be transferred between the solids, see Fig. \ref{asperity.eps}.
We assume the heat transfer occur in the area of real contact and that the heat current is proportional to
$\sigma ({\bf x})$, and to the temperature difference across the interface. The flow of heat energy from the
substrate to the rubber block can be described as an additional heat source acting on the surface $z=0$ of the rubber block:
$$\dot q_{\rm R} ({\bf x},z) = \alpha (T_{\rm S}-T_{\rm R}) \sigma({\bf x}-{\bf v}t)\delta (z)\eqno(33)$$
where $T_{\rm R}$ and $T_{\rm S}$ are the (average) rubber and substrate surface temperature in the contact regions.
It is often assumed that the temperature is continuous at the interface, which correspond to the limit $\alpha \rightarrow \infty$. 
Energy conservation imply that there will also be a heat source 
$$\dot q_{\rm S} ({\bf x},z) = -\alpha (T_{\rm S}-T_{\rm R}) \sigma({\bf x})\delta (z)\eqno(34)$$
on the substrate surface. Note that if the substrate is colder than the rubber
then $T_{\rm S}-T_{\rm R} <0$ and heat will flow from the rubber to the substrate. 

The contribution to the flash temperature in the rubber from $\dot q_{\rm R}$ is given by (30) with $\mu_{\rm ad} v$ replaced by $\alpha (T_{\rm S}-T_{\rm R})$.
Hence the heat transfer contributions to the flash temperature is
$$T_q ({\rm heat \ transfer}) = {\alpha (T_{\rm S}-T_{\rm R}) \over  2\pi \kappa \sigma_0} \int_{q_0}^{q_{1}} dk \, k^3 C(k) W(k)\int_0^{2\pi} d\phi \ 
 \int_{-\infty}^{\infty} dk_z {q |E^*(vk {\rm cos}\phi , T_k)|^2  \over (2q-i k_z)[k^2+k_z^2 - i (kv/D) {\rm cos}\phi ]}. \eqno(35)$$
The total flash temperature
$$T_q = T_q ({\rm visc})+T_q ({\rm ad})+T_q ({\rm heat \ transfer})$$
is the sum of the viscoelastic, adhesion and heat transfer contribution given by (27), (30) and (35), respectively.

The (average) surface temperature in the contact regions on the rubber side is
$$T_{\rm R} = T_{\rm R0}+T_\infty ({\rm visc})+T_\infty ({\rm ad})+T_\infty ({\rm heat \ transfer}), \eqno(36)$$
where $T_{\rm R0}$ is the rubber background temperature.

The flash temperature on the substrate is obtained from (30) with $\mu_{\rm ad} v$ replaced with $-\alpha (T_{\rm S}-T_{\rm R})$
and with $v=0$ in the denominator, 
and $\kappa$ replaced by the heat conductivity $\kappa_{\rm S}$ of the substrate. 
The surface ($z=0$) temperature corresponds to the $q\rightarrow \infty$ and can be obtained from (32)
$$T_{\rm S} = T_{\rm S0} -{ \alpha (T_{\rm S}-T_{\rm R})] \over  \kappa_{\rm S} \sigma_0} \int_{q_0}^{q_{
1}} dk \, k^2 C(k) W(k)\int_0^{\pi/2} d\phi \ |E^*(vk {\rm cos}\phi , T_k)|^2. \eqno(37)$$
where $T_{\rm S0}$ the substrate background temperature. 

Equation (36) and (37) are two equations for $T_{\rm R}$ and $T_{\rm S}$ which can be easily solved. 
Substituting the resulting $T_{\rm S}$ and $T_{\rm R}$ into (35), and adding (27) and (30) to get the total
$T_q$, result in an integral equation for $T_q$ (because $T_q$ inter in the viscoelastic modulus), 
which can be solved by iteration.

As stated above, in most cases one can assume that the temperature is continuous at the rubber-substrate interface.
This is the case if $\alpha$ is chosen large enough. However, in some cases this may not hold accurately. In particular
if the band of phonon frequencies of the substrate and the rubber differ a lot, the phonons may be reflected at the
interface, and the heat transfer ineffective \cite{PV}. This phonon mismatch thermal boundary resistance is often called the
Kapitza resistance, and was first experimentally observed by Kapitza during studies of heat transfer between solids and liquid helium. 

\vskip 0.2cm
{\bf 4 Summary and conclusions}

We have presented an analytical theory for the flash temperature for viscoelastic solids sliding on substrates 
with substrate surface roughness on arbitrary many decades in length scale.
The flash temperature is defined as a weighted average temperature where the weighting factor $\sigma ({\bf x})$
is proportional to the heat source. This definition of the flash temperature is the optimum average which can be defined.

The theory presented above can be extended to non-stationary (accelerated) sliding following the procedure reported
on in Ref. \cite{ToBe}. Using the same
approach as in Ref. \cite{M1} the theory can also extended to gives information about the 
variation of the temperature within the macroasperity contact regions,
as contained in derivatives of the temperature profile.
These generalizations, and numerical results illustrating the theory, 
will be presented elsewhere \cite{ToBe1}.

\end{document}